\documentclass[preprint,12pt]{elsarticle}

\usepackage[T1]{fontenc}
\usepackage[utf8]{inputenc}
\usepackage{lmodern}
\usepackage{amssymb}
\usepackage{amsmath}
\usepackage{booktabs}
\usepackage{tabularx}
\usepackage{threeparttable}
\usepackage{microtype}
\usepackage{xcolor}
\usepackage[hidelinks]{hyperref}

\graphicspath{{./}}
\biboptions{sort&compress}
\newcommand{\eLiMnOtwo}{\ensuremath{\varepsilon\text{-}\mathrm{LiMnO_2}}}
\newcommand{\LiMnOtwo}{\ensuremath{\mathrm{LiMnO_2}}}
\newcommand{\Liion}{\ensuremath{\mathrm{Li^+}}}
\newcommand{\paperfigure}[2][]{%
  \IfFileExists{#2}{\includegraphics[#1]{#2}}{%
    \fbox{\parbox[c][45mm][c]{0.90\linewidth}{\centering
    Figure file \texttt{#2} was not found.\\
    Place it in the same directory as this \texttt{.tex} file.}}}}

\begin{document}

\begin{frontmatter}

\title{Local coordination and migration-network topology shape Li-ion transport and delithiation in the low-energy \texorpdfstring{\eLiMnOtwo}{epsilon-LiMnO2} polymorph}

\author[ysu]{Fukuan Wang}
\author[ysu]{Busheng Wang}
\author[ysu]{Yong Liu\corref{cor1}}
\cortext[cor1]{Corresponding author.}
\ead{yongliu@ysu.edu.cn}

\affiliation[ysu]{organization={State Key Laboratory of Metastable Materials Science and Technology and Hebei Key Laboratory of Microstructural Material Physics, School of Science, Yanshan University},
                  city={Qinhuangdao},
                  postcode={066004},
                  state={Hebei},
                  country={China}}

\begin{abstract} In rocksalt-derived oxide cathodes, the local Li-migration environment around an O$_4$ tetrahedral intermediate is commonly classified by the number of face-sharing transition-metal (TM) neighbors. In LiMnO$_2$, the TM species is Mn, and 0-TM denotes the absence of face-sharing Mn neighbors. However, migration and delithiation may also depend on higher-shell coordination and tetrahedral connectivity. Using the recently reported low-energy $\varepsilon$-LiMnO$_2$ polymorph as a model, we examine these factors through bond-valence site-energy and bond-valence pathway analyses combined with first-principles calculations. The resulting migration maps and tetrahedral statistics reveal distinct topologies across four LiMnO$_2$ polymorphs. Although the $\varepsilon$ phase and the lithiated-spinel phase Li$_2$Mn$_2$O$_4$ (hereafter spinel) have identical tetrahedral-type fractions, their 0-TM motifs form quasi-one-dimensional chains and a three-dimensional network, respectively. Climbing-image nudged elastic band calculations yield $\varepsilon$-phase barriers of 0.35--0.36~eV, compared with 0.41--0.53~eV in spinel, a difference that may be associated with distinct next-nearest corner-sharing shells. Ab initio molecular dynamics yields an apparent activation energy of 0.32~eV, while direction-resolved mean-squared displacements show preferential Li migration along $c$, supporting low-barrier quasi-one-dimensional diffusion. Delithiation calculations further show that differences in 0-TM connectivity and Li--Li separation between the $\varepsilon$ phase and spinel are associated with Li-site evolution and calculated voltage steps. These results link local environments and the spatial connectivity of 0-TM motifs to Li migration and delithiation, providing a structural perspective for metastable cathode design. \end{abstract}

\begin{graphicalabstract}
\centering
\paperfigure[width=\linewidth]{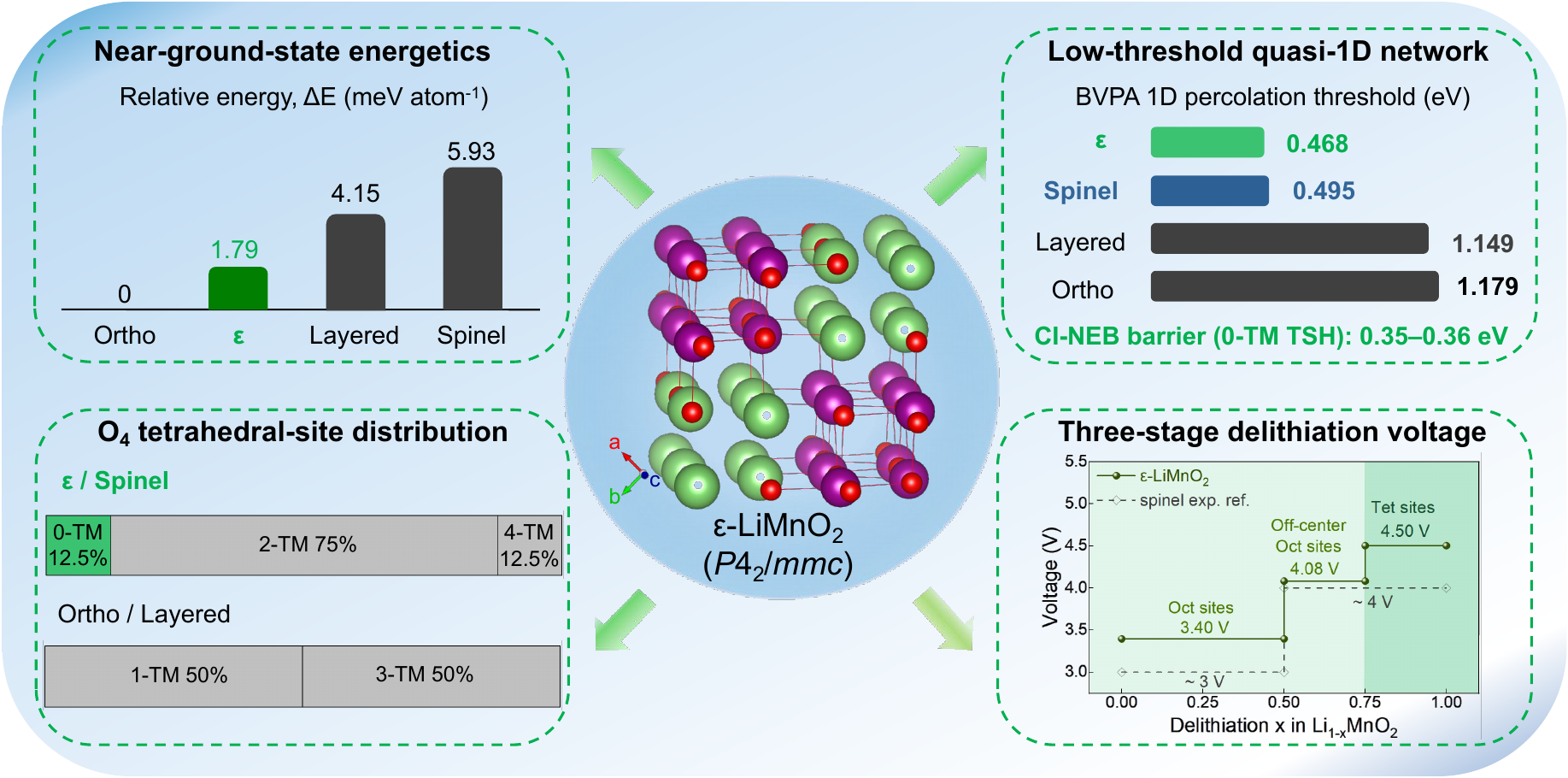}
\end{graphicalabstract}

\begin{highlights}
	\item $\varepsilon$-LiMnO$_2$ lies only 1.79~meV\,atom$^{-1}$ above orthorhombic LiMnO$_2$.
	\item Identical O$_4$-site statistics yield quasi-1D $\varepsilon$ and 3D spinel networks.
	\item Lower higher-shell MnO$_6$ density correlates with 0.35--0.36~eV 0-TM barriers.
	\item AIMD shows anisotropic Li diffusion with an activation energy of 0.32~eV.
	\item 0-TM connectivity regulates Li-site evolution and delithiation voltage.
\end{highlights}


\end{frontmatter}

%\linenumbers

\section{Introduction}

Growing demand for lithium-ion batteries across portable electronics, electric transportation, and grid-scale storage motivates cathodes that balance energy density, cost, safety, manufacturability, and resource sustainability. \cite{li30YearsLithiumIon2018,xieRetrospectiveLithiumionBatteries2020,turcheniukTenYearsLeft2018,huangKeyChallengesGridScale2022,liMaterialsCellStateoftheArt2022,xiaoLaboratoryInnovationsMaterials2023,tianPromisesChallengesNextgeneration2021,wuGuidelinesTrendsNextgeneration2020,costaRecyclingEnvironmentalIssues2021,wangDirectRegenerationSpent2024} Because cathodes largely determine cell energy density and contribute substantially to material cost, their crystal chemistry is central to battery design. \cite{manthiramReflectionLithiumionBattery2020,murdockPerspectiveSustainabilityCathode2021} Major cathode families---layered transition-metal oxides, spinels, and olivines---offer distinct tradeoffs in voltage, stability, resource availability, and processability. \cite{liHighnickelLayeredOxide2020,zhengChallengesAdvancementsHighnickel2026,xiangBuildingPracticalHighVoltage2022,zhuHighvoltageSpinelCathode2024,nekahiSustainableLiFePO4LiMnxFe1xPO42024}

Cost, concentrated supply chains, and sustainability concerns surrounding Co and Ni have renewed interest in Mn-rich and Ni/Co-free cathodes. \cite{liCobaltLithiumionBatteries2020,leeCanCobaltBe2022,liuUnderstandingCoRoles2021} Mn is abundant and inexpensive, while Mn-based disordered-rocksalt and layered oxides offer high capacity and structural tunability. \cite{liHighenergyMnbasedDisorderedrocksalt2022,parkZerostrainMnrichLayered2025} \LiMnOtwo\ is therefore attractive both as a high-energy cathode and as a model for cation-order effects \cite{miyaokaPracticalSustainableNi2024}, with a theoretical capacity of approximately 285~mAh\,g$^{-1}$. \cite{maProgressChallengeProspect2024} Its polymorphs include orthorhombic \LiMnOtwo\,\cite{dittrichZurKristallstrukturLiMnO21969}  layered \LiMnOtwo\, \cite{armstrongSynthesisLayeredLiMnO21996} and the $I4_1/amd$ lithiated-spinel phase Li$_2$Mn$_2$O$_4$ (hereafter spinel \LiMnOtwo). \cite{davidStructureRefinementSpinelrelated1987,kamInterplayElectronLocalization2025} Their behavior is strongly cation-order dependent: Jahn--Teller-active high-spin Mn$^{3+}$ is associated with MnO$_6$ distortions and may promote structural degradation, \cite{mishraStructuralStabilityLithium1999} whereas controlled cation disorder may improve reversibility. \cite{leeStructuralDisorderLayered2024}

In the established $n$-TM notation, $n$ counts the face-sharing
transition-metal cations around this tetrahedron. For the \LiMnOtwo\
polymorphs considered here, TM refers specifically to Mn. Low-$n$
environments, particularly 0-TM sites, therefore reduce Li--Mn repulsion
and favor migration through the tetrahedral intermediate.
\cite{leeUnlockingPotentialCationDisordered2014,clementCationdisorderedRocksaltTransition2020} This local descriptor is powerful, but incomplete. The fraction of favorable motifs does not specify whether they assemble into a continuous migration network, whereas network dimensionality and alternative-path availability are critical to long-range transport. \cite{urbanConfigurationalSpaceRocksaltType2014,clementCationdisorderedRocksaltTransition2020} Moreover, the same first-shell $n$-TM label contains no information about the higher-shell cation environment or about how neighboring tetrahedra are spatially connected. This distinction may matter not only for migration barriers but also during delithiation, when tetrahedral positions can evolve from transient migration intermediates into occupied Li sites, as occurs in the spinel framework. \cite{davidStructureRefinementSpinelrelated1987,goodenoughChallengesRechargeableLi2010} These considerations motivate examining whether structural information beyond the first-shell $n$-TM label---particularly higher-shell cation environments and tetrahedral connectivity---can help rationalize differences in Li migration behavior and, during delithiation, in Li-site evolution and the accompanying calculated voltage response.

Kam et al.\ recently identified an experimentally unreported \eLiMnOtwo\ structure with space group $P4_2/mmc$ through a cluster-expansion exploration of Li/Mn orderings and assessed its energetics using several electronic-structure treatments. \cite{kamInterplayElectronLocalization2025} Its collinearly ordered Jahn--Teller axes distinguish it from polymorphs with mixed or noncollinear distortion patterns, while its energy remains close to that of orthorhombic \LiMnOtwo\ and below those of the layered and lithiated-spinel phases across the treatments considered. The distinctive cation order of the $\varepsilon$ phase makes it an especially useful system for testing whether favorable local tetrahedral environments necessarily translate into connected transport and analogous delithiation behavior. Existing work, however, established its structural origin and low-energy character without addressing its dynamical, mechanical, and finite-temperature structural stability, Li migration topology and energetics, finite-temperature diffusion, or Li-site and voltage evolution upon delithiation.

Here, we establish the low-energy character and structural stability of \eLiMnOtwo\ and use it to examine how local tetrahedral environments and their spatial organization relate to Li migration and delithiation. A four-polymorph comparison maps their migration topologies, whereas a focused comparison of the $\varepsilon$ and spinel phases separates motif abundance from connectivity and probes possible higher-shell effects on migration barriers. Delithiation calculations further explore how 0-TM connectivity relates to Li-site evolution and the calculated voltage response. This local-to-global perspective provides a structural framework for evaluating metastable Mn-based rocksalt cathodes.

\section{Computational methods}

Density functional theory (DFT) calculations were performed using the Vienna
\textit{Ab initio} Simulation Package (VASP).
\cite{kresseEfficiencyAbinitioTotal1996,kresseEfficientIterativeSchemes1996}
The projector augmented-wave method
\cite{blochlProjectorAugmentedwaveMethod1994,
	kresseUltrasoftPseudopotentialsProjector1999} and the PBEsol functional
\cite{perdewGeneralizedGradientApproximation1996,
	perdewRestoringDensitygradientExpansion2008} were employed with a plane-wave
cutoff energy of 520~eV. Brillouin-zone integrations used Monkhorst--Pack
meshes with a reciprocal-space spacing of approximately 0.03~\AA$^{-1}$.
\cite{monkhorstSpecialPointsBrillouinzone1976} Structures were relaxed until
the total-energy change and residual atomic forces were below $10^{-5}$~eV
and 0.01~eV\,\AA$^{-1}$, respectively. The Mn $3d$ states were treated using
the Dudarev DFT+$U$ formalism with $U_{\mathrm{eff}}=3.9$~eV.
\cite{dudarevElectronenergylossSpectraStructural1998,
	zhouFirstprinciplesPredictionRedox2004,
	wangOxidationEnergiesTransition2006,jadidiEffectFluorinationLiexcess2020a}
The antiferromagnetic ground states reported by Kam et al.\ were adopted for
orthorhombic, layered, and spinel \LiMnOtwo{}. \cite{kamInterplayElectronLocalization2025} For \eLiMnOtwo, several
antiferromagnetic and ferromagnetic configurations were tested, and the
lowest-energy configuration was used subsequently (Fig.~S1).

Dynamical stability was evaluated by finite-displacement phonon calculations
using Phonopy.
\cite{bornDynamicalTheoryCrystal1996,togoFirstprinciplesPhononCalculations2023,
	togoImplementationStrategiesPhonopy2023} Thermal stability was examined by a
10~ps NVT AIMD simulation at 500~K using a 128-atom supercell, a
Nos\'e--Hoover thermostat, and a 2~fs time step.
\cite{noseUnifiedFormulationConstant1984,hooverCanonicalDynamicsEquilibrium1985}
Mechanical stability was assessed from the elastic constants using the
appropriate Born criteria. \cite{mouhatNecessarySufficientElastic2014}
Powder X-ray diffraction (XRD) patterns were simulated from the optimized structures using VESTA. \cite{mommaVESTA3Threedimensional2011}

Candidate Li sites and migration-network dimensionalities were screened using
bond-valence site-energy (BVSE) calculations in softBV and the Bond Valence
Pathway Analyzer (BVPA).
\cite{chenBondSoftnessSensitive2017,chenSoftBVSoftwareTool2019,
	wongBondValencePathway2021} Connectivity thresholds were referenced to the
lowest BVSE site in each structure using an energy cutoff of 2.5~eV.
Quantitative migration barriers were calculated in 128-atom supercells using
the climbing-image nudged elastic band (CI-NEB) method.
\cite{henkelmanClimbingImageNudged2000}
Finite-temperature Li-ion diffusion in \eLiMnOtwo\ was investigated by NVT
AIMD using a model constructed by removing four Li atoms from a 128-atom
parent structure. Simulations were performed at 750, 1000, 1500, and 2000~K,
with approximately 4~ps of equilibration followed by a 20~ps production
trajectory at each temperature. The Li mean-squared displacement (MSD) was
calculated as \cite{heStatisticalVariancesDiffusional2018}
\begin{equation}
	\mathrm{MSD}_{\mathrm{Li}}(t)
	=
	\frac{1}{N_{\mathrm{Li}}}
	\sum_{i=1}^{N_{\mathrm{Li}}}
	\left|
	\mathbf{r}_{i}(t)-\mathbf{r}_{i}(0)
	\right|^{2},
	\label{eq:msd}
\end{equation}
where $N_{\mathrm{Li}}$ is the number of Li ions. Direction-resolved MSDs were
obtained from the displacement components along the crystallographic $a$,
$b$, and $c$ directions. The diffusion coefficient was extracted from the
linear region of the total MSD using the Einstein relation
\cite{einsteinUberMolekularkinetischenTheorie1905}
\begin{equation}
	D
	=
	\frac{1}{6}
	\frac{\mathrm{d}\,\mathrm{MSD}_{\mathrm{Li}}(t)}
	{\mathrm{d}t}.
	\label{eq:einstein}
\end{equation}
The diffusion coefficients obtained at 1000, 1500, and 2000~K were fitted
using the Arrhenius relation
\cite{arrheniusUberReaktionsgeschwindigkeitBei1889}
\begin{equation}
	D(T)
	=
	D_{0}
	\exp\left(
	-\frac{E_{\mathrm{a}}}{k_{\mathrm{B}}T}
	\right),
	\label{eq:arrhenius}
\end{equation}
where $D_{0}$ is the pre-exponential factor, $k_{\mathrm{B}}$ is the
Boltzmann constant, and $E_{\mathrm{a}}$ is the apparent diffusion activation
energy.

Li/vacancy orderings on the parent octahedral sublattice were enumerated using the Supercell program. \cite{okhotnikovSupercellProgramCombinatorial2016} At each Li content, the ten symmetry-inequivalent configurations with the lowest electrostatic energies were relaxed, together with selected tetrahedral-site configurations motivated by the known octahedral-to-tetrahedral Li rearrangement in spinel during delithiation. \cite{davidStructureRefinementSpinelrelated1987,goodenoughChallengesRechargeableLi2010} Their energies were used to construct the configurational convex hull of $\mathrm{Li}_{1-x}\mathrm{MnO_2}$, where $x$ denotes the delithiation fraction.
The formation energy was calculated as
\cite{ongLiFePO2PhaseDiagram2008,urbanComputationalUnderstandingLiion2016}
\begin{equation}
	\Delta E_{\mathrm f}(x)
	=
	E(\mathrm{Li}_{1-x}\mathrm{MnO_2})
	-(1-x)E(\mathrm{LiMnO_2})
	-xE(\mathrm{MnO_2}).
	\label{eq:formation}
\end{equation}
For delithiation from $x_1$ to $x_2$ ($x_2>x_1$), the average open-circuit
voltage relative to Li/Li$^+$ was calculated as
\cite{urbanComputationalUnderstandingLiion2016,
	aydinolInitioStudyLithium1997}
\begin{equation}
	V(x_1\rightarrow x_2)
	=
	-\frac{
		E(\mathrm{Li}_{1-x_1}\mathrm{MnO_2})
		-E(\mathrm{Li}_{1-x_2}\mathrm{MnO_2})
		-(x_2-x_1)E(\mathrm{Li}_{\mathrm{bcc}})
	}{
		x_2-x_1
	},
	\label{eq:voltage}
\end{equation}
where $E(\mathrm{Li}_{\mathrm{bcc}})$ is the energy per atom of bulk bcc Li.

\section{Results and discussion}

\subsection{Structure and stability of \texorpdfstring{\eLiMnOtwo}{epsilon-LiMnO2}}

\eLiMnOtwo\ crystallizes in space group $P4_2/mmc$, with Li and Mn occupying the $4k$ and $4l$ Wyckoff sites, respectively, and O occupying the $4j$ and $4m$ sites. The two inequivalent O sites are each coordinated by six cations drawn from the Li and Mn sublattices. The oxygen sublattice retains a cubic-close-packed arrangement similar to those of other \LiMnOtwo\ polymorphs, such that the $\varepsilon$ phase can be viewed as an ordered Li/Mn derivative of the rocksalt framework. Among the AFM1, AFM2, AFM3, and ferromagnetic initial states tested for this structure, AFM1 had the lowest energy, consistent with the magnetic ground state reported by Kam et al.{}. \cite{kamInterplayElectronLocalization2025} The complete comparison is provided in Fig.~S1.

\begin{figure}[htbp]
	\centering
	\paperfigure[width=\linewidth]{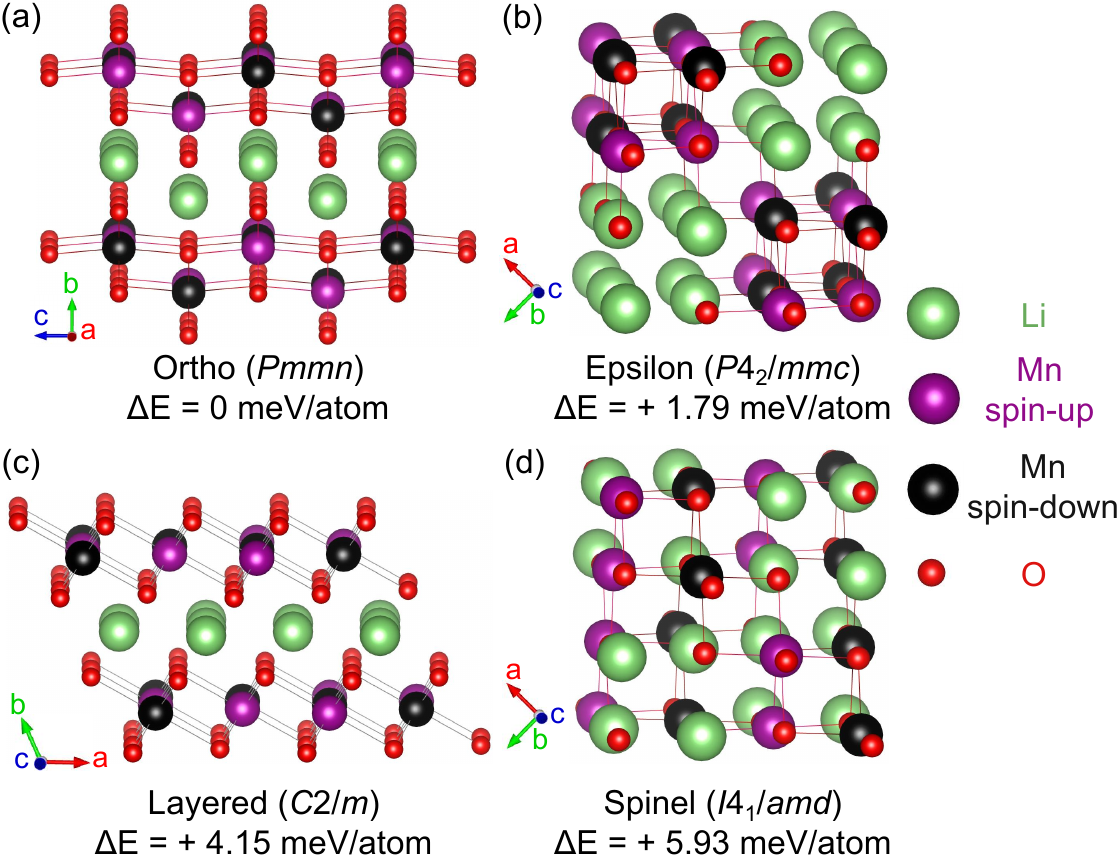}
	\caption{Optimized structures and relative energies of four \LiMnOtwo\ polymorphs: (a) orthorhombic (denoted as Ortho), (b) epsilon ($\varepsilon$), (c) layered, and (d) spinel. The spinel polymorph denotes the lithiated $I4_1/amd$ structure, equivalent to Li$_2$Mn$_2$O$_4$ in conventional spinel notation. Energies are referenced to orthorhombic \LiMnOtwo. Green, purple, black, and red spheres represent Li, spin-up Mn, spin-down Mn, and O, respectively.}
	\label{fig:polymorphs}
\end{figure}

Figure~\ref{fig:polymorphs} compares the relaxed structures and energies of the four polymorphs. \eLiMnOtwo\ is 1.79~meV\,atom$^{-1}$ above orthorhombic \LiMnOtwo, whereas the layered and spinel structures are 4.15 and 5.93~meV\,atom$^{-1}$ above the orthorhombic reference, respectively. The small energy separation places the $\varepsilon$ phase among the low-energy competing polymorphs of \LiMnOtwo. Optimized lattice parameters and fractional coordinates are given in Tables~S1 and S2.

\begin{figure}[htbp]
	\centering
	\paperfigure[width=\linewidth]{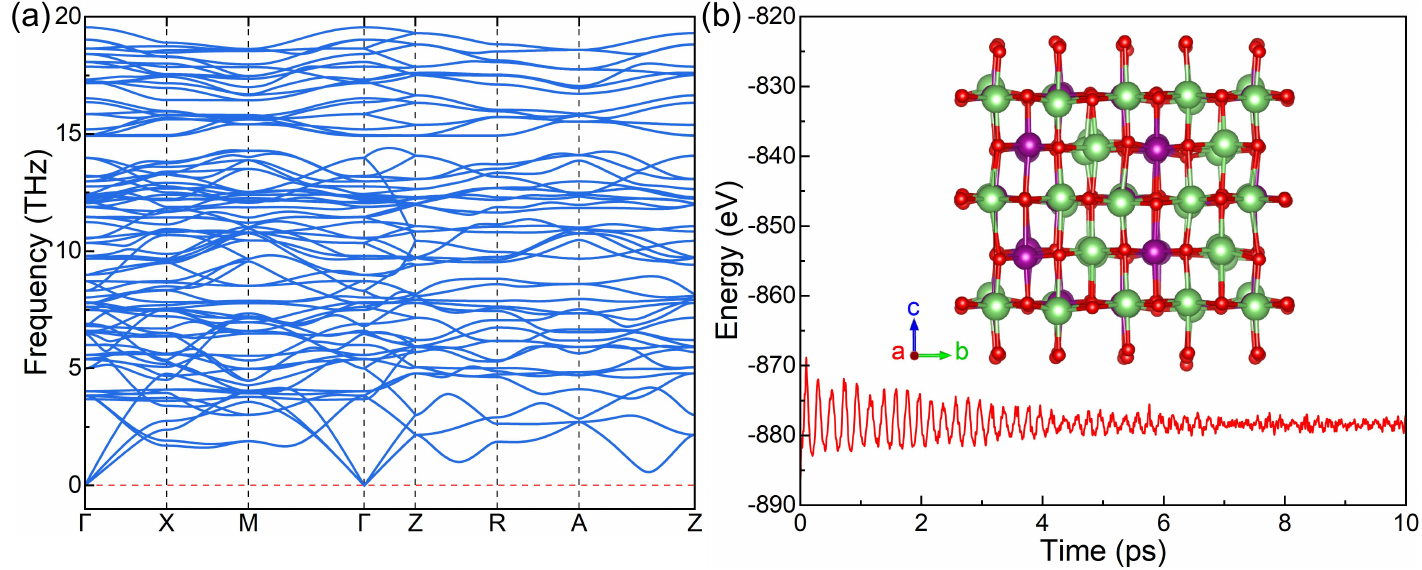}
	\caption{Dynamical and finite-temperature structural stability of \eLiMnOtwo. (a) Phonon dispersion, showing no imaginary modes. (b) Total energy during a 10~ps AIMD trajectory at 500~K; the inset shows the final configuration, which retains the original structural framework.}
	\label{fig:stability}
\end{figure}

The phonon dispersion in Fig.~\ref{fig:stability}(a) contains no imaginary modes, indicating that the optimized structure is dynamically stable. During the 500~K AIMD simulation [Fig.~\ref{fig:stability}(b)], the total energy fluctuates within a bounded range and the framework remains intact over the 10~ps trajectory, indicating structural stability at the simulated temperature. The calculated elastic tensor satisfies the Born mechanical-stability criteria for a tetragonal crystal (Table~S3), and the corresponding Voigt--Reuss--Hill elastic properties are summarized in Table~S4. Taken together with its low relative energy, these results motivate examination of the electrochemical behavior of \eLiMnOtwo.

Simulated powder X-ray diffraction patterns provide a structural fingerprint for distinguishing the polymorphs (Fig.~S2). The $\varepsilon$ and orthorhombic phases display similar low-angle features, including reflections near $15$--$16^\circ$ and $25^\circ$. By contrast, the strongest low-angle reflections of the layered and spinel phases occur near $18$--$19^\circ$, and neither shows the intense feature near $25^\circ$. The close resemblance between the $\varepsilon$ and orthorhombic patterns may hinder their distinction in experimental powder XRD data, particularly in the presence of peak broadening or phase overlap, and could explain why the $\varepsilon$ phase has remained unidentified in previously reported \LiMnOtwo\ samples. \cite{kamInterplayElectronLocalization2025}

\subsection{Migration topology and O\texorpdfstring{$_4$}{4} environments}

The dimensionality of the \Liion\ migration network is an important structural factor governing long-range ionic transport in electrode materials. Higher-dimensional networks generally provide more interconnected and alternative migration pathways and are less susceptible to local channel blocking, which can facilitate macroscopic Li transport and rate performance. \cite{clementCationdisorderedRocksaltTransition2020,urbanConfigurationalSpaceRocksaltType2014} On this basis, BVSE and BVPA were employed to compare the connectivity and dimensionality of the \Liion\ migration networks in the four \LiMnOtwo\ polymorphs.

\begin{figure}[htbp]
	\centering
	\paperfigure[width=\linewidth]{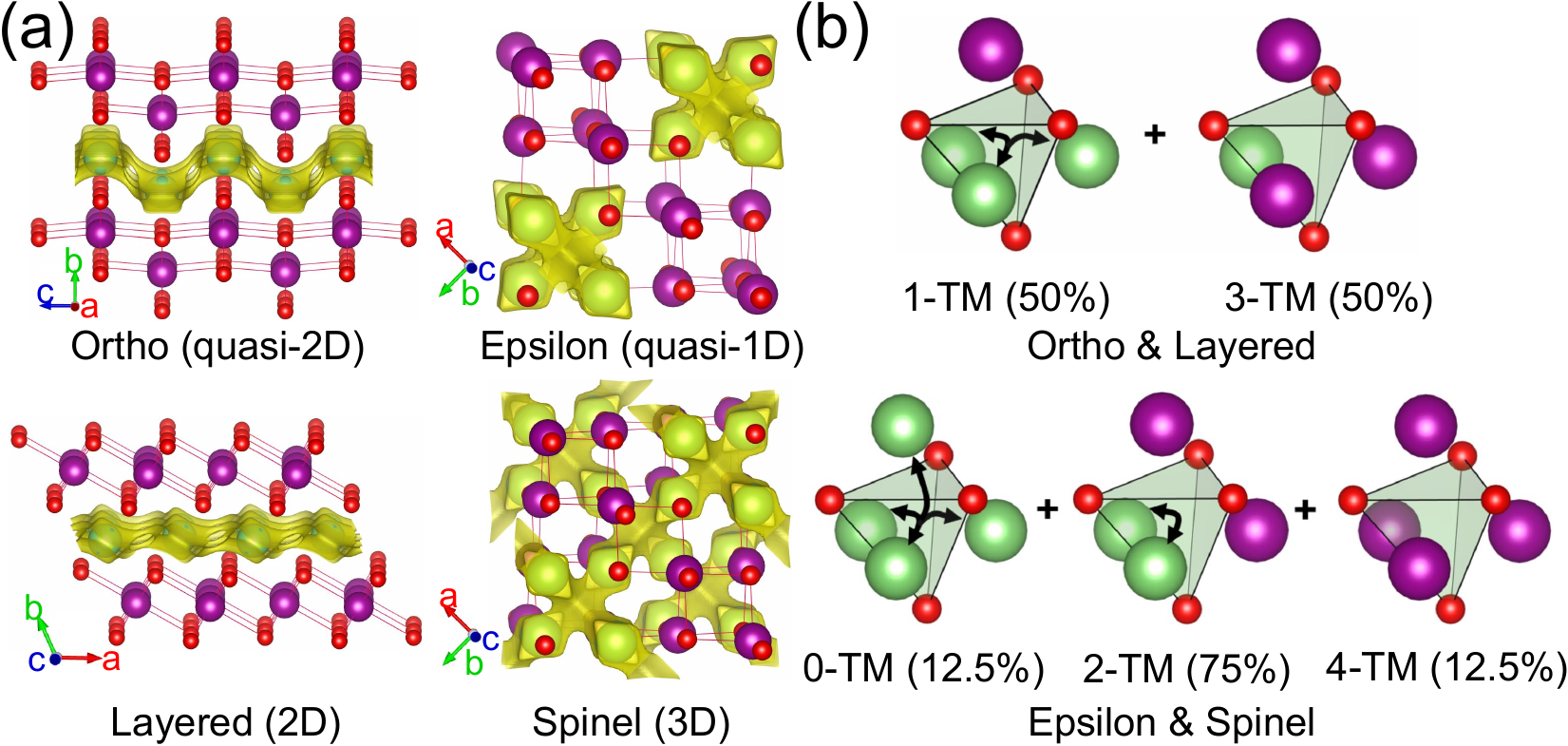}
	\caption{Migration topology and O$_4$ tetrahedral environments in four \LiMnOtwo\ polymorphs. (a) BVSE isosurfaces superimposed on the Mn/O frameworks. Yellow isosurfaces denote low-BVSE regions for \Liion, and the same isosurface level of $-1$ was used for all four structures. Orthorhombic, $\varepsilon$, layered, and spinel \LiMnOtwo\ exhibit quasi-two-dimensional, quasi-one-dimensional, two-dimensional, and three-dimensional migration networks, respectively. (b) Distribution of O$_4$ tetrahedral sites classified according to their local TM environments. The orthorhombic and layered phases each contain 50\% 1-TM and 50\% 3-TM sites, whereas the $\varepsilon$ and spinel phases each contain 12.5\% 0-TM, 75\% 2-TM, and 12.5\% 4-TM sites.}
	\label{fig:bvse}
\end{figure}

The BVSE landscapes reveal distinct spatial organizations of the low-energy \Liion\ regions [Fig.~\ref{fig:bvse}(a)]. In orthorhombic \LiMnOtwo, these regions connect mainly along two directions, forming a quasi-two-dimensional network. In \eLiMnOtwo, the low-energy regions extend primarily along one crystallographic direction and form separated quasi-one-dimensional chains. Layered \LiMnOtwo\ exhibits a two-dimensional network within the Li layers, whereas spinel \LiMnOtwo\ forms a three-dimensionally interconnected network.

The $n$-TM distributions in Fig.~\ref{fig:bvse}(b) separate the four
polymorphs into two pairs: the orthorhombic and layered structures each
contain 50\% 1-TM and 50\% 3-TM sites, whereas the $\varepsilon$ and spinel structures each contain 12.5\% 0-TM, 75\% 2-TM, and 12.5\% 4-TM sites. Nevertheless, the low-energy network in the spinel is connected in three dimensions, whereas the corresponding regions in the $\varepsilon$ phase are connected primarily along quasi-one-dimensional chains. The dimensionality of the migration network is therefore determined not only by the statistical distribution of local TM environments but also by the spatial arrangement and connectivity of the tetrahedral sites.

\begin{table}[!t]
	\caption{BVPA connectivity thresholds and inferred \Liion\ migration topology in \LiMnOtwo\ polymorphs}
	\label{tab:bvpa}
	\begin{threeparttable}
		\begin{tabular*}{\columnwidth}
			{@{\extracolsep{\fill}}lcccl@{}}
			\toprule
			Structure & 1D (eV) & 2D (eV) & 3D (eV) & Topology \\
			\midrule
			Ortho   & 1.179 & 1.329 & N.C.\tnote{a} & quasi-2D \\
			Epsilon & 0.468 & 2.031 & 2.031          & quasi-1D \\
			Layered & 1.149 & 1.347 & N.C.\tnote{a} & 2D \\
			Spinel  & 0.495 & 0.495 & 0.495          & 3D \\
			\bottomrule
		\end{tabular*}
		\begin{tablenotes}[flushleft]
			\footnotesize
			\item[a] N.C.: no percolating network of the specified dimensionality was identified within 2.5~eV of the lowest BVSE site of the corresponding structure.
		\end{tablenotes}
	\end{threeparttable}
\end{table}

The BVPA connectivity thresholds in Table~\ref{tab:bvpa} further quantify these topological differences. Orthorhombic \LiMnOtwo\ reaches one- and two-dimensional connectivity at 1.179 and 1.329~eV, respectively, whereas no three-dimensional network is identified within the search range. The $\varepsilon$ phase reaches one-dimensional connectivity at a relatively low threshold of 0.468~eV, while the formation of two- and three-dimensional networks requires a much higher threshold of 2.031~eV, indicating that its low-energy migration network is dominated by the quasi-one-dimensional direction. Layered \LiMnOtwo\ reaches one- and two-dimensional connectivity at 1.149 and 1.347~eV, respectively, without forming a three-dimensional network within the search range. In spinel \LiMnOtwo, one-, two-, and three-dimensional connectivity all emerge at 0.495~eV, consistent with its fully three-dimensional network.

Taken together, the BVSE/BVPA maps and O$_4$ $n$-TM statistics distinguish two complementary structural levels of Li-ion migration: the first-shell environment of each tetrahedral intermediate and the spatial connection of the corresponding O--T--O segments into extended pathways. The contrast between the $\varepsilon$ and spinel phases shows that identical local motif fractions do not necessarily yield the same migration-network topology, demonstrating that the first-shell $n$-TM classification alone is insufficient to infer transport dimensionality.

\subsection{Local migration pathways and activation barriers}

\begin{figure*}[htbp]
	\centering
	\paperfigure[width=\linewidth]{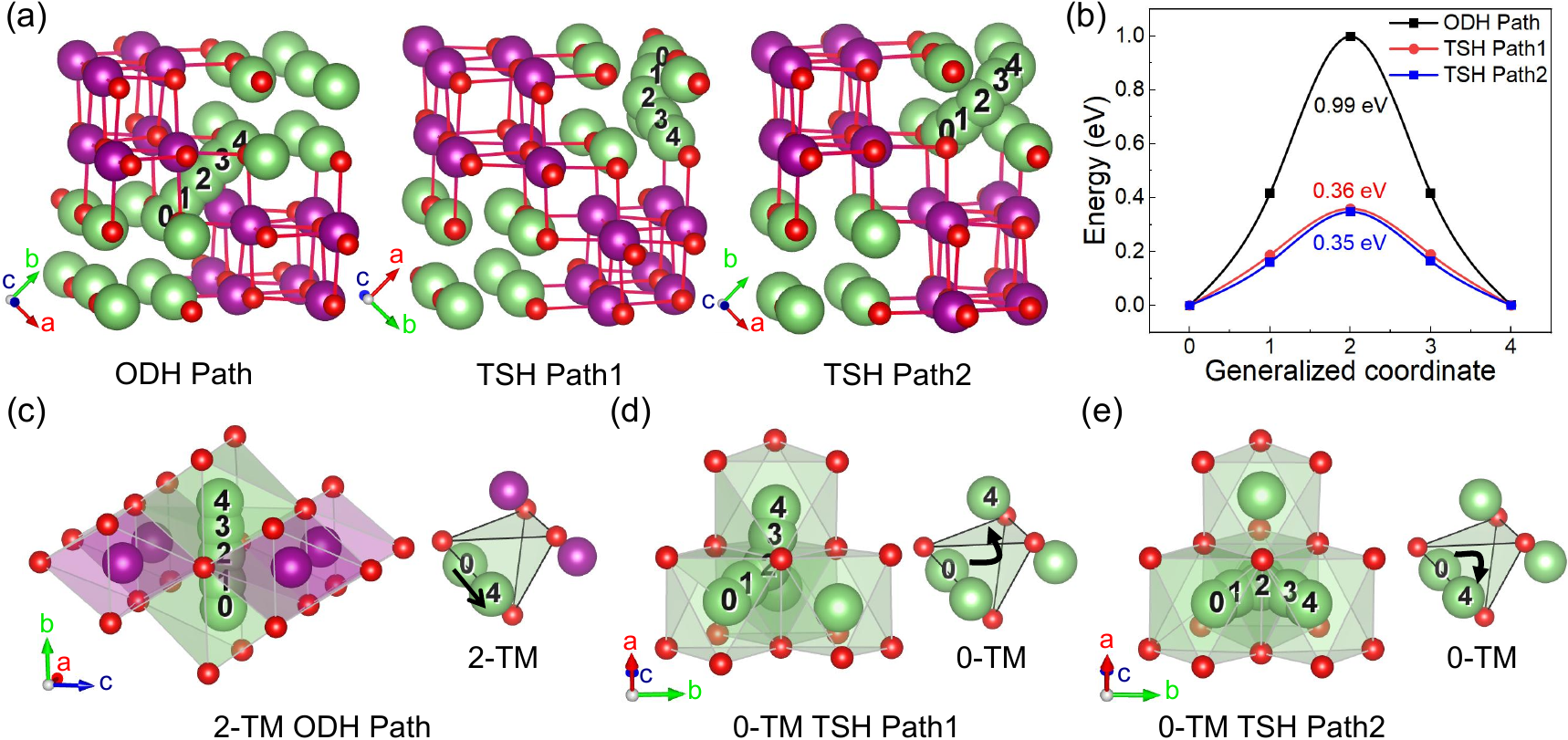}
	\caption{\Liion\ migration pathways and their local structural environments in \eLiMnOtwo\ under the single-vacancy model. (a) Structural representations of the ODH pathway and TSH pathways 1 and 2. (b) Corresponding CI-NEB energy profiles. (c--e) Local tetrahedral environments associated with the 2-TM ODH pathway, 0-TM TSH pathway 1, and 0-TM TSH pathway 2, respectively. Black arrows indicate the migration direction, and labels 0--4 mark successive positions of the migrating Li along each pathway. Green, purple, and red spheres represent Li, Mn, and O, respectively.}
	\label{fig:epsilon-neb}
\end{figure*}

Based on the BVSE/BVPA results, three representative single-vacancy \Liion\ hops in \eLiMnOtwo\ were examined using CI-NEB, with the corresponding 0-TM hops in the lithiated-spinel phase included for comparison. In rocksalt-derived oxides, migration between neighboring octahedral sites can proceed through either an oxygen-dumbbell hop (ODH), in which Li passes through an O--O dumbbell bottleneck, or a tetrahedral-site hop (TSH) through an O$_4$ interstice. \cite{vandervenLithiumDiffusionLayered1999,vandervenLithiumDiffusionMechanisms2001} When the tetrahedral intermediate is energetically unfavorable, the migrating Li tends to deviate from the tetrahedral center and follow an ODH trajectory. By contrast, a low-TM tetrahedral environment can stabilize the intermediate and enable a lower-barrier TSH process.

Figure~\ref{fig:epsilon-neb}(a,b) shows one ODH pathway and two TSH pathways in \eLiMnOtwo. Their calculated migration barriers are 0.99, 0.36, and 0.35~eV, respectively, demonstrating the strong dependence of local \Liion\ migration on the surrounding cation environment.

The high-barrier pathway is associated with a 2-TM tetrahedral environment [Fig.~\ref{fig:epsilon-neb}(c)]. A Li ion passing through the corresponding tetrahedral center would face-share with two MnO$_6$ octahedra and experience strong Li--Mn repulsion. During CI-NEB relaxation, the migrating Li therefore avoids the high-energy tetrahedral center and follows an ODH trajectory close to the O--O dumbbell bottleneck. The resulting barrier of 0.99~eV indicates that this 2-TM connection is unlikely to contribute significantly to the low-energy migration network.

By contrast, both low-barrier pathways pass through 0-TM tetrahedral sites [Fig.~\ref{fig:epsilon-neb}(d,e)]. The absence of face-sharing MnO$_6$ octahedra reduces Li--Mn repulsion and allows the tetrahedral position to serve as a relatively stable migration intermediate. The resulting barriers of 0.36 and 0.35~eV are consistent with the local O$_4$ classification and the low one-dimensional BVPA connectivity threshold. However, these favorable local hops do not establish isotropic diffusion. The 0-TM TSH pathways are aligned mainly along the quasi-one-dimensional chains, whereas transverse transport is likely constrained by high-barrier connections such as the examined 2-TM pathway. Consequently, \eLiMnOtwo\ combines low local migration barriers with restricted global connectivity.

\begin{table}[!t]
	\caption{Single-vacancy \Liion\ migration mechanisms and barriers in \LiMnOtwo\ polymorphs}
	\label{tab:barriers}
	\begin{tabular*}{\columnwidth}
		{@{\extracolsep{\fill}}llcl@{}}
		\toprule
		Structure & Path type & Barrier (eV) & Source \\
		\midrule
		Ortho   & ODH & 0.50 / 0.56 & Ref.~\cite{jadidiEffectFluorinationLiexcess2020a} \\
		Epsilon & TSH & 0.36 / 0.35 & This work \\
		Layered & ODH & 0.58 / 0.63 & Refs.~\cite{hoangDefectPhysicsDelithiation2015,kongInitioStudyDoping2015} \\
		Spinel  & TSH & 0.41 / 0.53 & This work \\
		\bottomrule
	\end{tabular*}
\end{table}

\begin{figure*}[htbp]
	\centering
	\paperfigure[width=\linewidth]{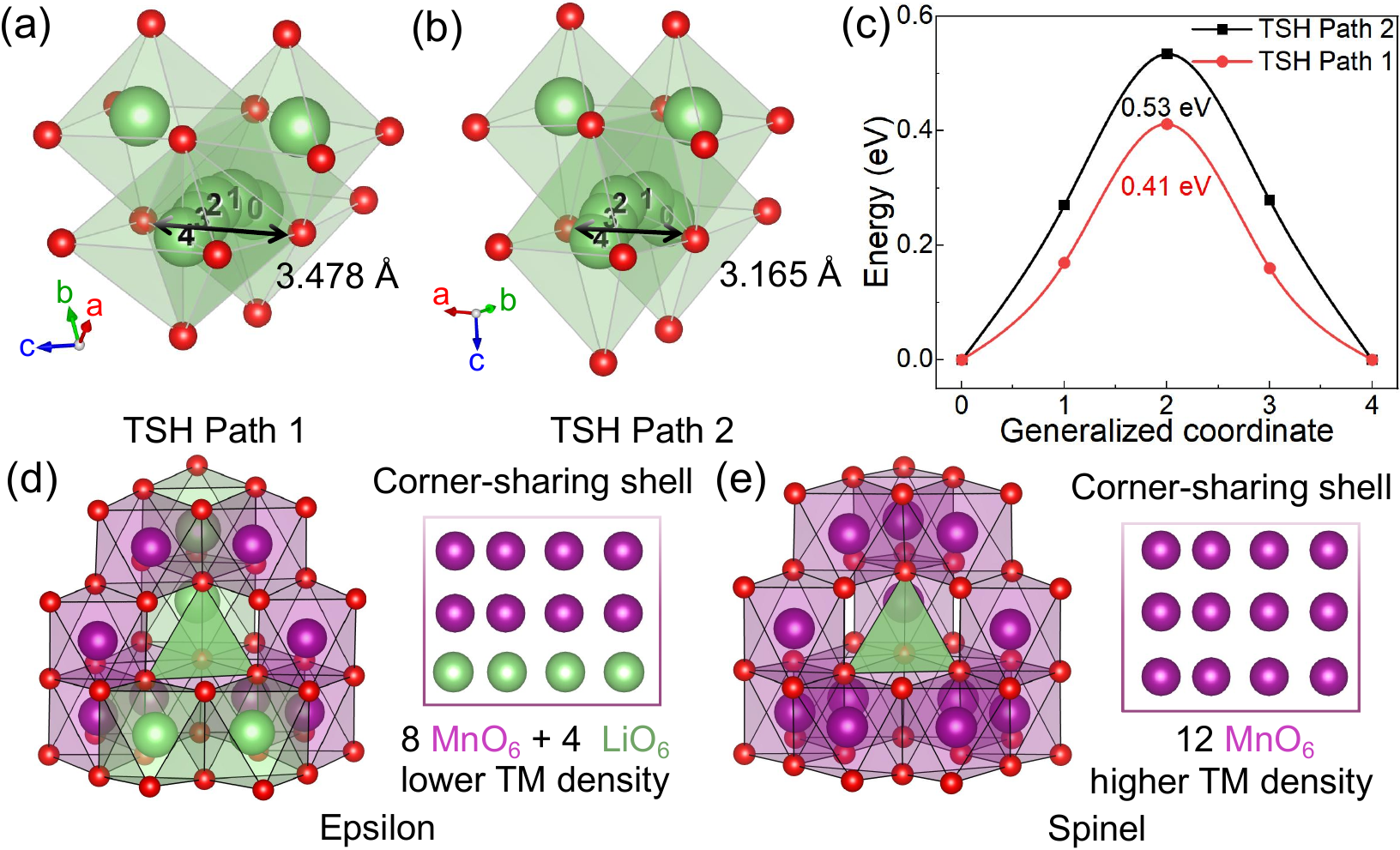}
	\caption{0-TM TSH migration in spinel \LiMnOtwo\ and comparison of its local cation environment with that of \eLiMnOtwo. (a,b) Local geometries of TSH pathways 1 and 2 in the spinel, with O--O bottleneck widths of 3.478 and 3.165~\AA, respectively. (c) Corresponding CI-NEB energy profiles, with migration barriers of 0.41 and 0.53~eV. The higher barrier of pathway~2 is associated with its narrower O--O bottleneck. (d,e) Corner-sharing octahedral shells surrounding a 0-TM tetrahedral site in \eLiMnOtwo\ and spinel \LiMnOtwo, respectively. The shell contains eight MnO$_6$ and four LiO$_6$ octahedra in the $\varepsilon$ phase, whereas all twelve octahedra are MnO$_6$ in the spinel. Black arrows indicate Li migration, and labels 0--4 mark successive positions of the migrating Li. Green, purple, and red spheres represent Li, Mn, and O, respectively.}
	\label{fig:spinel-neb}
\end{figure*}

Table~\ref{tab:barriers} compares the calculated barriers with those reported for other \LiMnOtwo\ polymorphs. Previous calculations reported ODH barriers of 0.58/0.63~eV for layered \LiMnOtwo\ and 0.50/0.56~eV for orthorhombic \LiMnOtwo{}. \cite{hoangDefectPhysicsDelithiation2015,kongInitioStudyDoping2015,jadidiEffectFluorinationLiexcess2020a} The two 0-TM TSH barriers in the $\varepsilon$ phase are lower than these literature values. For comparison with another 0-TM TSH system, two inequivalent pathways were also calculated for spinel \LiMnOtwo. As shown in Fig.~\ref{fig:spinel-neb}(a--c), their migration barriers are 0.41 and 0.53~eV, with corresponding O--O bottleneck widths of 3.478 and 3.165~\AA\ for pathways~1 and 2, respectively. Within the spinel structure, the narrower O--O bottleneck is associated with the higher migration barrier, indicating that local bottleneck geometry can differentiate migration kinetics even when both pathways traverse 0-TM tetrahedral sites.

Although the migration pathways in both structures have the same first-shell 0-TM environment, the two barriers in the $\varepsilon$ phase (0.36 and 0.35~eV) are lower than those in the spinel. To examine a possible structural origin of this difference, the next-nearest corner-sharing octahedral shell surrounding the migration-related 0-TM tetrahedron was compared [Fig.~\ref{fig:spinel-neb}(d,e)]. In \eLiMnOtwo, this shell consists of eight MnO$_6$ and four LiO$_6$ octahedra, whereas all twelve octahedra are MnO$_6$ in spinel \LiMnOtwo. The lower Mn density in this shell may reduce longer-range Li--Mn repulsion and structural constraints near the tetrahedral intermediate, thereby contributing to the lower 0-TM TSH migration barriers in the $\varepsilon$ phase.

\begin{figure}[!t]
	\centering
	\paperfigure[width=\linewidth]{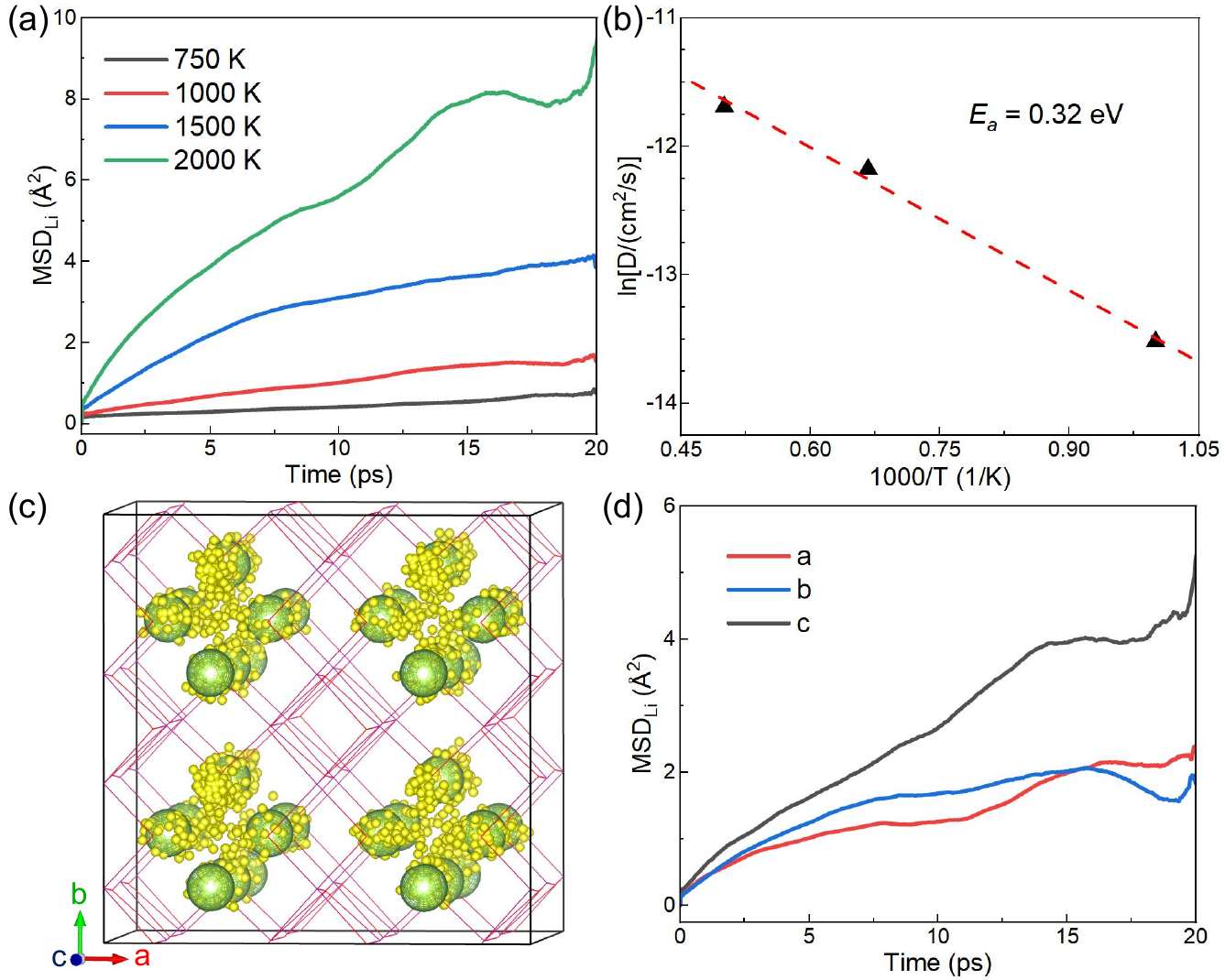}
	\caption{Finite-temperature \Liion\ diffusion in \eLiMnOtwo\ obtained from AIMD simulations. (a) Li mean-squared displacements (MSDs) at 750, 1000, 1500, and 2000~K, shown by gray, red, blue, and green curves, respectively. (b) Arrhenius plot based on the diffusion coefficients obtained at 1000--2000~K. Black triangles denote the calculated values, and the red dashed line denotes the linear fit, yielding an apparent diffusion activation energy of 0.32~eV with a coefficient of determination of $R^2=0.9944$. (c) Superimposed instantaneous Li positions sampled every 40 frames from the 2000~K AIMD trajectory. For clarity, the Mn--O host framework is displayed in wireframe mode. Green spheres indicate the initial Li positions, whereas yellow spheres denote the sampled Li positions. (d) Direction-resolved Li MSDs along the crystallographic $a$, $b$, and $c$ directions at 2000~K, shown by red, blue, and gray curves, respectively.}
	\label{fig:aimd}
\end{figure}

Building on the CI-NEB analysis, AIMD simulations were further performed to examine finite-temperature \Liion\ diffusion in \eLiMnOtwo. The temperature-dependent MSDs, Arrhenius fitting, Li-ion trajectories, and direction-resolved MSDs are summarized in Fig.~\ref{fig:aimd}.

As shown in Fig.~\ref{fig:aimd}(a), the Li MSD increases markedly with temperature, indicating thermally activated \Liion\ migration in \eLiMnOtwo. At 750~K, the Li MSD remains relatively small, and sustained long-range diffusion is not observed within the limited simulation time of 20~ps. As the temperature increases to 1000, 1500, and 2000~K, the MSD grows progressively faster, indicating that elevated temperatures increase the probability of Li ions overcoming local migration barriers and undergoing successive hopping events. The element-resolved MSDs further show that Mn and O remain largely confined to thermal vibrations around their equilibrium positions at all four temperatures, whereas the pronounced displacement originates primarily from Li ions. This result indicates that the Mn--O host framework remains stable during the AIMD simulations (Fig.~S3).

The diffusion coefficients at 1000--2000~K were extracted from the linear regions of the MSD curves according to Eq.~\ref{eq:einstein}, and the apparent diffusion activation energy was subsequently determined by Arrhenius fitting according to Eq.~\ref{eq:arrhenius}. As shown in Fig.~\ref{fig:aimd}(b), $\ln D$ exhibits a linear dependence on $1000/T$, yielding an apparent diffusion activation energy of 0.32~eV for \eLiMnOtwo.
This value is close to the CI-NEB barriers of 0.35 and 0.36~eV for the two 0-TM TSH pathways, indicating good consistency between the \Liion\ migration energy scales obtained from the two methods.

Figure~\ref{fig:aimd}(c) further shows the sampled instantaneous Li positions from the 2000~K AIMD trajectory. Rather than being uniformly distributed throughout the crystal framework, the Li positions form continuously extended migration regions primarily along the $c$ direction, revealing pronounced spatial anisotropy in the finite-temperature Li motion. The direction-resolved MSDs provide quantitative support for this trajectory distribution. As shown in Fig.~\ref{fig:aimd}(d), at approximately 20~ps and 2000~K, the Li MSDs along the $c$, $a$, and $b$ directions are approximately 5.2, 2.3, and 1.9~\AA$^2$, respectively. Although appreciable Li motion also occurs within the $ab$ plane, the MSD along $c$ is more than twice the corresponding values along $a$ and $b$, demonstrating preferential \Liion\ migration along the $c$ direction and further supporting the quasi-one-dimensional migration channels identified above.

A hierarchical picture of structural control over \Liion\ migration therefore emerges: the number of first-shell face-sharing Mn neighbors primarily determines the migration mechanism and characteristic barrier scale, whereas the MnO$_6$ density in the next-nearest corner-sharing shell may provide a weaker modulation among pathways with the same $n$-TM classification. Within this framework, \eLiMnOtwo\ uniquely combines low-barrier TSH hopping with a quasi-one-dimensional migration network. Such restricted connectivity may hinder efficient long-range transport despite favorable local hopping kinetics, potentially limiting the practical rate capability.

\subsection{Delithiation voltage and evolution of Li coordination}

\begin{figure*}[htbp]
	\centering
	\paperfigure[width=\linewidth]{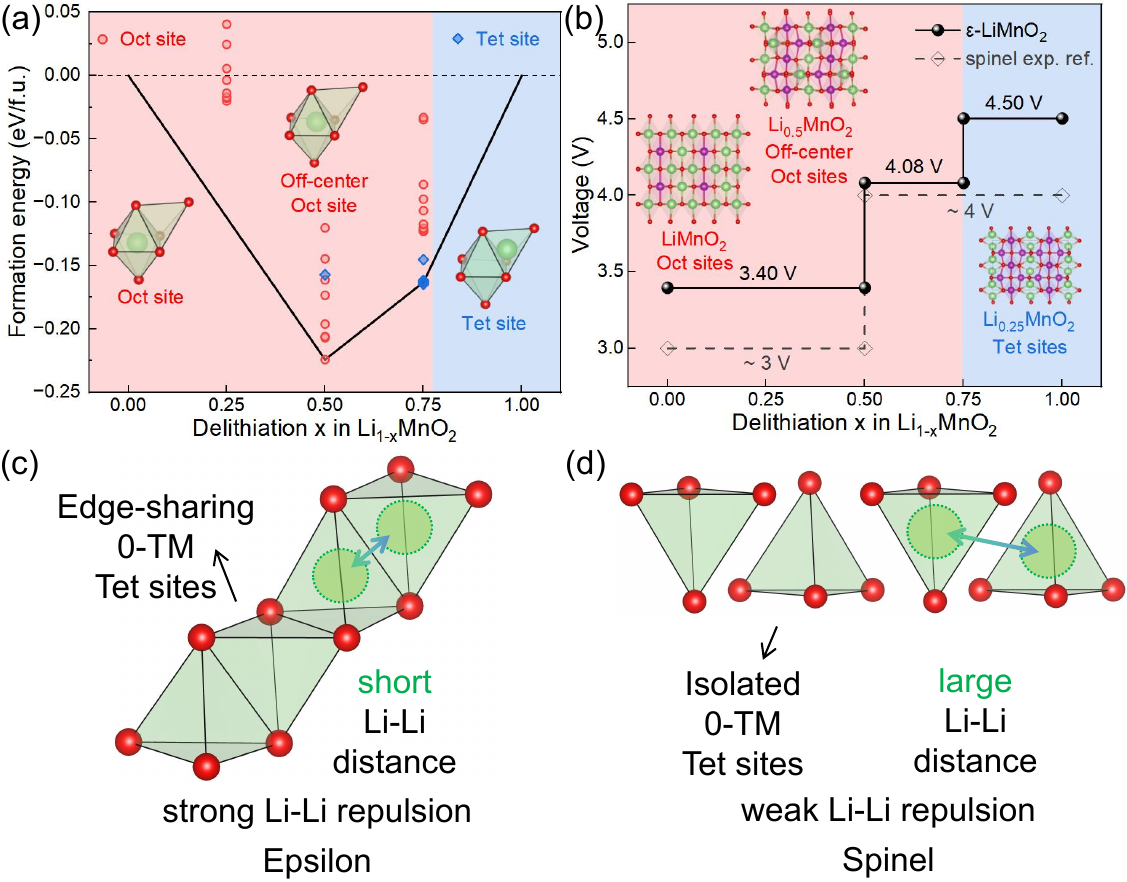}
	\caption{Formation energies, voltage evolution, and local Li-site environments during delithiation of \eLiMnOtwo. (a) Formation energies of candidate $\mathrm{Li}_{1-x}\mathrm{MnO_2}$ configurations. Red circles and blue diamonds denote configurations initialized with predominantly octahedral and tetrahedral Li occupation, respectively. The black line denotes the sampled configurational convex hull. (b) Calculated average delithiation voltages. Gray dashed lines show the representative experimental voltage plateaus of spinel Li$_z$Mn$_2$O$_4$.\cite{goodenoughChallengesRechargeableLi2010,thackerayElectrochemicalExtractionLithium1984} The spinel compositions are normalized per Mn such that Li$_2$Mn$_2$O$_4$, LiMn$_2$O$_4$, and Mn$_2$O$_4$ correspond to LiMnO$_2$, Li$_{0.5}$MnO$_2$, and MnO$_2$, respectively. The background colors indicate the different Li-site regimes in the $\varepsilon$ phase. (c) Edge-sharing 0-TM LiO$_4$ tetrahedral sites in a tetrahedral candidate configuration of $\varepsilon$-Li$_{0.5}$MnO$_2$. (d) Relatively isolated 0-TM tetrahedral sites in spinel Li$_{0.5}$MnO$_2$.}
	\label{fig:delithiation}
\end{figure*}

The tetrahedral sites discussed above can serve as intermediate sites for Li migration and may also become occupied during delithiation. To examine how local motif connectivity relates to Li-site evolution and voltage response in \eLiMnOtwo, the formation energies of candidate $\mathrm{Li}_{1-x}\mathrm{MnO_2}$ configurations and the average delithiation voltages between adjacent low-energy compositions were calculated, and the evolution of the local Li environment was analyzed. As shown in Fig.~\ref{fig:delithiation}(a), at $x=0$, the LiO$_6$ coordination cage is already distorted because of the Jahn--Teller distortion of the surrounding Mn$^{3+}$O$_6$ octahedra and the associated cooperative lattice response. \cite{marianettiFirstprinciplesInvestigationCooperative2001,kamInterplayElectronLocalization2025} The six Li--O distances are $2\times2.037$, $2\times2.132$, and $2\times2.275$~\AA. Despite this distortion, Li remains at the center of the distorted O$_6$ cage. At $x=0.25$, corresponding to Li$_{0.75}$MnO$_2$, all sampled configurations lie above the tie line connecting the adjacent low-energy compositions, and no intermediate vertex therefore appears on the sampled convex hull. By contrast, low-energy intermediate configurations occur at $x=0.50$ and $x=0.75$, corresponding to Li$_{0.5}$MnO$_2$ and Li$_{0.25}$MnO$_2$, respectively.

At Li$_{0.5}$MnO$_2$, Li in the lowest-energy configuration no longer occupies the center of the O$_6$ coordination cage but exhibits a pronounced off-center displacement. The local environment contains four shorter Li--O bonds, $2\times1.939$ and $2\times2.006$~\AA, together with two longer Li--O contacts of $2\times2.587$~\AA. This position is therefore described as an off-center octahedral site, distinguishing it from the centered Li position in the distorted octahedron of the fully lithiated structure. Upon further delithiation, Li occupies a tetrahedral site in the sampled low-energy Li$_{0.25}$MnO$_2$ configuration, with four Li--O distances of 1.952--2.030~\AA. Overall, the sampled low-energy configurations indicate that Li occupation evolves from the center of a distorted octahedron, through an off-center position within the O$_6$ cage, and eventually to a tetrahedral site. Correspondingly, the delithiation voltages are 3.40, 4.08, and 4.50~V, associated with centered octahedral occupation, off-center octahedral occupation, and tetrahedral occupation at lower Li contents, respectively.

For comparison, Fig.~\ref{fig:delithiation}(b) also shows the classical experimental voltage plateaus of spinel Li$_z$Mn$_2$O$_4$ near 3 and 4~V. \cite{davidStructureRefinementSpinelrelated1987,thackerayElectrochemicalExtractionLithium1984,goodenoughChallengesRechargeableLi2010} The approximately 3~V process corresponds to the Li$_2$Mn$_2$O$_4$/LiMn$_2$O$_4$ interval and involves the removal of octahedral Li, whereas the approximately 4~V process corresponds to the removal of tetrahedral Li from LiMn$_2$O$_4$ toward $\lambda$-MnO$_2$. After normalization per Mn, tetrahedral Li occupation is already stabilized in spinel Li$_{0.5}$MnO$_2$, whereas the $\varepsilon$ phase retains off-center octahedral occupation at the same composition and evolves toward tetrahedral coordination only in the sampled configurations at lower Li contents. Accordingly, the 3.40~V segment of the $\varepsilon$ phase lies above the approximately 3~V spinel plateau, the 4.08~V segment is close to the approximately 4~V plateau, and the voltage further increases to 4.50~V upon deeper delithiation. The $\varepsilon$ phase therefore exhibits a three-stage voltage response distinct from the conventional two-plateau behavior of the spinel.

To clarify the structural origin of the different Li occupations despite the identical fraction of 0-TM tetrahedral sites, the spatial connectivity of these sites was compared between the $\varepsilon$ and spinel phases [Fig.~\ref{fig:delithiation}(c,d)]. In the tetrahedral candidate configuration of $\varepsilon$-Li$_{0.5}$MnO$_2$, neighboring 0-TM LiO$_4$ tetrahedra share edges, resulting in a nearest-neighbor Li--Li distance of only 2.150~\AA. This distance is shorter than the 2.562~\AA\ separation in the lowest-energy off-center-octahedral configuration, suggesting that stronger local Li--Li repulsion may occur in the tetrahedral candidate configuration. By contrast, the 0-TM tetrahedral sites in spinel Li$_{0.5}$MnO$_2$ are relatively isolated, with a nearest-neighbor Li--Li distance of 3.589~\AA; this larger separation is consistent with weaker local Li--Li repulsion and stable tetrahedral occupation. These results indicate that, in addition to the fraction of 0-TM sites, their spatial connectivity and the associated Li--Li separations are also related to Li-site preferences and the accompanying calculated voltage response during delithiation.

\section{Conclusions}

This work identifies \eLiMnOtwo\ as a low-energy and structurally stable \LiMnOtwo\ polymorph and, through comparison among four polymorphs, reveals a hierarchical structural control over \Liion\ migration. The number of first-shell face-sharing Mn neighbors primarily governs the migration mechanism and characteristic barrier scale, whereas the MnO$_6$ density in the next-nearest corner-sharing shell may provide weaker secondary modulation. The spatial connectivity of migration segments passing through 0-TM tetrahedral intermediates determines the directionality and dimensionality of the long-range migration network. Although the $\varepsilon$ and lithiated-spinel phases contain identical fractions of O$_4$ tetrahedral-site types, they form quasi-one-dimensional and three-dimensional low-energy networks, respectively, demonstrating that $n$-TM statistics alone cannot predict long-range transport characteristics.

Within this framework, \eLiMnOtwo\ combines low-barrier 0-TM TSH hopping with quasi-one-dimensional long-range migration. The CI-NEB barriers of 0.35--0.36~eV agree with the apparent activation energy of 0.32~eV obtained from AIMD, while the direction-resolved MSDs confirm preferential \Liion\ diffusion along the crystallographic $c$ direction. However, the restricted quasi-one-dimensional network may impede efficient long-range transport and limit the practical rate capability. During delithiation, adjacent 0-TM tetrahedra are edge-sharing in the $\varepsilon$ phase rather than isolated as in the spinel. This connectivity increases Li--Li repulsion in candidate tetrahedral configurations, such that Li remains at off-center positions within O$_6$ coordination cages in $\varepsilon$-Li$_{0.5}$MnO$_2$, whereas tetrahedral occupation appears only in sampled low-energy $\varepsilon$-Li$_{0.25}$MnO$_2$ configurations upon further delithiation, accompanied by a distinct stepwise voltage response.

Experimental synthesis and electrochemical characterization are therefore needed to validate the phase stability, diffusion anisotropy, rate capability, and delithiation-induced structural evolution of \eLiMnOtwo. More broadly, the analytical framework developed here, which integrates local $n$-TM environments, higher-shell coordination, and global network topology, can be extended to other rocksalt-derived cathodes to clarify how different structural levels govern ionic migration and electrochemical evolution and to test the generality of this local-to-global picture.

\section*{CRediT authorship contribution statement}

\textbf{Fukuan Wang:} Methodology, Software, Validation, Formal analysis, Investigation, Data curation, Visualization, Writing -- original draft. \textbf{Busheng Wang:} Validation, Formal analysis, Investigation, Writing -- review and editing. \textbf{Yong Liu:} Conceptualization, Supervision, Project administration, Writing -- review and editing.

\section*{Declaration of competing interest}
The authors declare that they have no known competing financial interests or personal relationships that could have appeared to influence the work reported in this paper.

\section*{Acknowledgements}

This work was supported by the National Natural Science Foundation of China (Grant No. 12504025), the Hebei Natural Science Foundation
(Grant No. A2025203011), and the Innovation Capability Improvement Project of Hebei Province (Grant No. 22567605H). This work also received support from the French government program ``Investissements d'Avenir'' as part of the France 2030 initiative (EUR INTREE, ANR-18-EURE-0010). The numerical calculations in this paper were performed on the supercomputing system at the High Performance
Computing Center of Yanshan University.

\section*{Data availability}
The data that support the findings of this study are available from the
corresponding author upon reasonable request.

\bibliographystyle{elsarticle-num}
\bibliography{references}

\end{document}